\documentclass{article}

\usepackage[preprint, nonatbib]{neurips_2023}

\usepackage[utf8]{inputenc} 
\usepackage[T1]{fontenc}    
\usepackage{hyperref}       
\usepackage{url}            
\usepackage{booktabs}       
\usepackage{tabularx}
\usepackage{graphicx}
\usepackage{amsfonts}       
\usepackage{nicefrac}       
\usepackage{microtype}      
\usepackage{xcolor}         
\usepackage{amsmath}
\usepackage{enumitem} 
\setlist[itemize]{nosep,wide=0pt, leftmargin=*, before=\vspace{-0.5\baselineskip}, after=\vspace{-\baselineskip}}

\title{State Space Paradox of Computational Research in Creativity}

\author{%
  \"{O}mer Ak{\i}n\\
  Carnegie Mellon University\\
  Pittsburgh, PA 15213 \\
  \texttt{oa04@andrew.cmu.edu} \\
  \And
  Yuning Wu \\
  Carnegie Mellon University\\
  Pittsburgh, PA 15213 \\
  \texttt{yuningw@andrew.cmu.edu} \\
}

\begin{document}

\maketitle

\begin{abstract}
  This paper explores the paradoxical nature of computational creativity, focusing on the inherent limitations of closed digital systems in emulating the open-ended, dynamic process of human creativity. Through a comprehensive analysis, we delve into the concept of the State Space Paradox (SSP) in computational research on creativity, which arises from the attempt to model or replicate creative behaviors within the bounded state spaces of digital systems. Utilizing a combination of procedural and representational paradigms, we examine various computational models and their capabilities to assist or emulate the creative process. Our investigation encompasses rule-based systems, genetic algorithms, case-based reasoning, shape grammars, and data mining, among others, to understand how these methods contribute to or fall short of achieving genuine creativity. The discussion extends to the implications of SSP on the future of creativity-related computer systems, emphasizing the cultural and contextual fluidity of creativity itself and the challenges of producing truly creative outcomes within the constraints of pre-defined algorithmic structures. We argue that while digital systems can provoke sudden mental insights (SMIs) in human observers and potentially support the creative process, their capacity to autonomously break out of their pre-programmed state spaces and achieve originality akin to human creativity remains fundamentally constrained. The paper concludes with reflections on the future directions for research in computational creativity, suggesting that recognizing and embracing the limitations and potentials of digital systems could lead to more nuanced and effective tools for creative assistance.
\end{abstract}

\section{Computational Research in Creativity}

It is an often-used adage that humans are fundamentally curious and creative. Yet, some take issue with the implication that creativity is innate and argue that we gain power over goals through knowledge, whether they are related to creativity or not. This makes a case for a pragmatic view of our efforts to explore, inquire and research: “the human condition can be improved through understanding.” Ultimately, all explanations of the human drive to achieve novelty are based on the tautological notion that creativity and curiosity have value. Regardless of the motivations underlying it, understanding the phenomenon around us will, eventually, turn out to be important. Through such understanding, we recognize, describe, emulate and control external (i.e., global climate) as well as internal (i.e., human psyche) phenomena.

\subsection{Sudden Mental Insight: A Form of Creativity}
One of the widely recognized and studied forms of creative behavior is the one called the Sudden Mental Insight (SMI). This phenomenon has received considerable coverage in creativity literature \cite{hayes2013complete}. SMI refers to the sudden onset of a realization that makes the solution of a very difficult problem or the creation of a remarkable result possible. Hayes \cite{hayes2013complete} argues that SMIs can be explained through already-known cognitive functions. Others have shown how the creative “leap” is akin to bridge building between the problem and solution domains which are normally separated by a chasm, and described the mechanics of the SMI in the context of several design and problem-solving protocols. While, to date, important issues remain unresolved and un-researched, SMI is one of the few, known, overt signs of creativity \cite{akin2020creativity}.

\subsection{Creativity and Computation}
In the age of Information Technology (IT), it is rare that any topic should be untouched by tools and concepts of computation. It turns out that creativity is one of the earliest IT goals addressed by techno-savvy folks of all kinds and backgrounds. It is no wonder that artist Harold Cohen has been painting with brush as well as Aaron, his digital counterpart, for more than three decades \cite{holtzman1995digital}. Cohen’s motivation for building the digital painter Aaron was for the same reasons as those provided in the introduction to this essay: curiosity and the impulse to do something new, which happens to be a curiously circular explanation. In the early years, Aaron was an automaton following instructions given to it through “rules,” a common device used in most Artificial Intelligence (AI) applications. Soon, Cohen realized that Aaron was no match for a human painter, like himself, principally because it did not learn from its experiences. Upon the urging of Edward Feigenbaum, who is considered to be one of the fathers of AI, Cohen decided to write some rules into Aaron about color theory. Then, Aaron started using color, which elicited the “wow!” or the SMI response from Cohen himself. He asks: “How did it come up with that?” providing us with a living example of how computer systems can behave in ways that are surprisingly human. Obviously, such personal impressions alone cannot be the measure of machine intelligence.

Alan Turing presented a test for machine intelligence through a succinct description. "I propose to consider the question, 'Can machines think?'" or "Are there imaginable digital computers which would do well in the imitation game?" \cite{turing2009computing} Ultimately this sort of thinking led to the following tangible proposition: “It is not difficult to devise a paper machine which will play a not very bad game of chess. Now get three men as subjects for experiments A, B, and C. A and C are to be rather poor chess players, and B is the operator who works the paper machine. ... Two rooms are used with some arrangement for communicating moves, and a game is played between C and either A or the paper machine. C may find it quite difficult to tell which he is playing.” \cite{turing2009computing} Hence, the general principle is that if we are unable to distinguish between a digital agent and a human by observing only their behavior – whether playing chess or reciting poetry – then we must consider the digital agent as capable as its human counterpart. Yet, Harold Cohen, like so many other users of digital assistants in creative tasks, considers these tools inferior because they can neither act in novel ways of their own volition nor learn from their actions.

Eve Sussman created a program with the help of Jeff Garneau, called the “Serendipity Machine,” which makes real-time splices of a set of video and audio recordings, based on a pre-defined, index-matching schema. As the permutations of audio-video pairings are spliced end-to-end the result turns out to be quite startling if not delightful. Yet, Sussman is unwilling to call the Serendipity Machine a “creative companion.” Professor Selmer Bringsjord of Rensselaer Polytechnic Institute believes that mystifying the creator of the digital system is the least a creativity system should do; otherwise, he concludes that “we will keep cloning our own intelligence.” Brigham Young University scientists have built a system called Darcy that judges artworks. Darcy has elicited curiosity among humans, yet upon learning that its judgment is based solely on a preference for “red, bloody, and violent”, our enthusiasm wanes. We have many digital emulators of human activities but lack the litmus test for what is sufficiently creative, or intelligent. Bringsjord brings this idea home when he remarks “Martha Stewart is credited with being creative when she recommends that we should use brown napkins with a yellow table cloth.” Up to now, the Turing test is the best thing anyone has come up with; yet, even that would not be able to show that airplanes are not as capable as birds, even though they can out-fly, out-distance, and out-cargo birds.

\section{Theoretical background and open-ended issues: Computer Assistance in Creativity}
Computer-based research on creativity, even from the beginning, has focused on a combination of procedural and representational paradigms. Digital system models of creativity, on the other hand, build models through a singular feature, either representational or procedural, but not both. Procedural approaches include (1) rule-based expert systems, (2) case-based reasoning systems, and (3) complex generative algorithms (such as genetic, annealing, and neural nets); while representational ones include: (1) shape emergence, (2) object-based representation, and (3) complex recognition systems (data mining, Petri-Nets).

\subsection{Procedural Approaches}
All software, regardless of its primary functionality must operate within a representation. Expert systems tend to use the rewrite-rule formalism for this purpose. Case-based reasoning approaches match, retrieve, and adapt cases to create new solutions. Genetic algorithms rely on the representation of complex symbolic strings called genotypes that can map into complex objects. Mimicking the lateral inhibitions that take place between the ganglia during synaptic activity in the cerebral cortex, neural nets are representations that are built in order to create lateral relations between the nodes of a network. While representation is important, essentially, these approaches are built to provide procedurally defined approaches it machine intelligence. Representations are there, merely to facilitate the procedural objectives by enabling genetic mutations, rule firings, case adaptations, or neural-net derivations that can achieve creative solutions (Table \ref{procedural}).

\begin{table}[htbp]
  \caption{Procedural Systems for Design Creativity}
  \label{procedural}
  \centering
  \resizebox{\textwidth}{!}{%
    \begin{tabularx}{\textwidth}{lXX}
      \toprule
      & \textbf{Procedural Schema} & \textbf{Representation Schema} \\
      \midrule
      Rule-based systems & Apply rewrite rules that have their left-hand side match problem representation & 
      \begin{itemize}
        \item Problem parameters-variables
        \item Rewrite rules
        \item Strategy for rule application
      \end{itemize} \\
      \midrule
      Genetic algorithms & Use meta-rules to mutate rewrite rules and generate solutions & 
      \begin{itemize}
        \item Problem parameters-variables
        \item Rewrite rules
        \item Strategy for rule application
        \item Rule mutation mechanism
      \end{itemize} \\
      \midrule
      Case-based systems & 
      \begin{itemize}
        \item Match case
        \item Retrieve case
        \item Adapt case
      \end{itemize} & 
      \begin{itemize}
        \item Case representation
        \item Case-based
      \end{itemize} \\
      \bottomrule
    \end{tabularx}%
  }
\end{table}

Several researchers have explored the potential of genetic algorithms in design. Often, the design domain is represented as a collection of rules. The mutation of these rules holds great promise in effecting change in design search space. Using a search metaphor to explore the design space and their genetic metamorphosis illustrates the power of such approaches. Difficulty, however, exists in the predictability of the results based on the modifications made to the rules.

Rule-based representations have given rise to the conjecture that design can be achieved through the application of predetermined rules of geometric composition. The potential of the approach has been amply demonstrated by many who have created design spaces after well-known, often historical sets and styles of designs: Palladian plans, Ire-Ray windows, and Queen Anne houses. A counter-intuitive but promising result that has emerged from the early work in this area is that the grammar formalism often goes far beyond the original set of patterns and designs that give rise to the grammar, in the first place.

Maher’s work on case-based engineering design demonstrates how precedents can be used to create paths of evolution for new designs starting from existing ones \cite{maher1997case}. Some may argue that creative solutions should not be based on precedents or cases. Others argue that all designs, novel or routine, are based on earlier examples. In the end, the adaptation functionality that transforms the case into a solution makes it possible to reach a non-routine, if not novel, design. In summary, the creative process envisioned by these systems requires that the problem being solved be represented in terms dictated by the procedural algorithm.

\subsection{Representational Approaches}
Because it is versatile enough to be regarded as a representational approach as well, Shape Grammar has been an important area of investigation in design creativity (Table \ref{representational}). This is largely due to their potential to recognize emergent shapes \cite{stiny2006shape}. In some cases, creativity is attributed to the ability of the designer to detect patterns that are not evident but are “evolving.” The quality of a design then is affected by these points of SMI that a designer recognizes as she is navigating in a space of design solutions.


\begin{table}[htbp]
  \caption{Representational Systems for Design Creativity}
  \label{representational}
  \centering
  \resizebox{\textwidth}{!}{%
    \begin{tabularx}{\textwidth}{XXX}
      \toprule
      & \textbf{Representation Schema} & \textbf{Procedural Schema} \\
      \midrule
      Shape emergence and grammars & 
      \begin{itemize}
        \item Geometric primitives
        \item Maximal shapes
      \end{itemize} & 
      Combinatorial enumeration \\
      \midrule
      Cognitive schema & 
      Object-based representation of functional, behavioral, and physical characteristics & 
      \begin{itemize}
        \item Formal reasoning
        \item Heuristic reasoning
      \end{itemize} \\
      \midrule
      Recognition algorithms (data mining, Petri-Nets) & 
      \begin{itemize}
        \item Large databases
        \item Process models
      \end{itemize} & 
      \begin{itemize}
        \item Pattern recognition
        \item Heuristic search
        \item Abstraction
      \end{itemize} \\
      \bottomrule
    \end{tabularx}%
}
\end{table}

Others argue that in order to represent the process of creativity a more complex representational schema is needed, including functions, behaviors, and structures to be embedded in new designs \cite{coyne2013computer}. This goes back to the early schema-based linguistic representations of memory and more recent applications in object-based software engineering approaches that have also been applied in architectural design. While these approaches also have great potential in capturing nontrivial aspects of architectural design, their claim of creativity has not been demonstrated \cite{rosenman1993creativity}.

In very complex design space networks, Petri-Nets, and colored Petri-Nets in particular, can abstract general patterns that are not evident to the naked eye. These applications are most useful in representing complex procedural domains, such as VLSI design or large system design problems in chemical plants. Through these applications, it is possible to control and predict overall performance in designed systems, including error detection and recovery, time of completion, and cost of delivery.

Data mining, a complex pattern recognition algorithm, is even more general in its purposes. It allows the user to discern patterns in unorganized data or data organized for purposes other than the ones currently at hand. Through this, it is possible to identify relevant design requirements or select among many alternative solutions the
ones that are most likely to yield creative solutions.

\section{Implications for Theory, Policy, and Practice: The Paradox of Creativity Research}
\subsection{The State Space of Creativity}
All digital systems of creativity, whether intended for assistance or emulation of the process, exist within an implicit or explicit state space \cite{simon1978information, newell1965simulation}. The state space represents any finite slice of time in the digital system’s functionality through entities, operations, goals, heuristics, and predicates that apply to that moment in time. This is a powerful concept because it enables us to talk about the digitally modeled process of creativity, or any formalized process, in discrete terms.

At any time slice, the digital application works with representational and procedural applications towards satisfying a goal (Tables \ref{procedural}, \ref{representational}, \ref{hybrid}). This goal may be to determine if a given object is creative (i.e., Darcy), or to create an object that emulates features we may consider creative (i.e., Aaron, Serendipity Machine). In either case, the details of the outcomes are computable from the specifics of the state space. All that goes into the computation, whether it is a set of criteria to interpret patterns and colors on a painting, rules of color theory, a generative algorithm to transform a given genotype, an emergent pattern, or the requirement specification for a layout generator, all is subsumed in the state space representation. In other words, these systems like all other computer programs are closed systems. Because their input parameters and possible outcomes are predefined, they cannot behave in any manner that is not pre-programmed through these definitions.

A human agent, on the other hand, is an open system and functions in an evolving state space. She modifies the initial state, the methods of operation that transform states, and the scope of acceptable solutions, at will \cite{simon1973structure}. In other words, depending on the circumstance she may prefer blue, sad and subdued over red, bloody, and violent; as well as to shift the criteria of selection to a voting mechanism by onlookers. The permutations are as endless as concepts carried in one’s head, including those that are not possible to express in words or represent in symbolic notation.

\subsection{Environments for Integrating Representations and Procedures}
Computational environments created to support the mixing and matching of representation with procedural formalisms can provide support for design creativity. The ingredients necessary for such integration are extremely demanding. Table \ref{hybrid} shows an illustrative scenario in which many representations and procedures can be used in tandem to reach creative solutions to a design problem. In such a scenario, the designer starts with an object-based representation, which allows her to reason about the overall behavior of the object to be designed, its functional characteristics, and its structure. The design proposal emerging from this can be used to search case-based legacy designs to see if similar solutions have been developed in the past and if the present solution can be improved using their features. Here, the designer may observe that a shape-grammatical order is evident. In that case, the Shape Grammar formalism can be used to detect pattern emergence; and Genetic Algorithms can be used to realize a design mutation suggested by the emergent patterns. Finally, the designer performs Data Mining to discern the dominant features of the solutions generated and represents these using the original schemata consisting of functions, behaviors, and structures.

\begin{table}
  \caption{Creative Search Scenario Based on a Hybrid Assistance System driven by Sudden Mental Insights (SMIs).}
  \label{hybrid}
  \centering
  \resizebox{\textwidth}{!}{%
  \begin{tabular}{lll}
    \toprule
    \textbf{Problem State} & \textbf{Example of problem reformulations driven by SMIs} & \textbf{Representation or Procedural System} \\
    \midrule
    T0 - Initial State & Cognitive schema based initial problem formulation & Cognitive schema \\
    T1 - 1\textsuperscript{st} SMI & Case-based solutions & Case-based system \\
    T2 - 2\textsuperscript{nd} SMI & Shape formalism rule-based solutions & Rule-based system \\
    T3 - 3\textsuperscript{rd} SMI & Emergent shape-based solutions & Shape grammar formalism \\
    T4 - 4\textsuperscript{th} SMI & Generative rule-based transformation of solutions & Genetic algorithm \\
    T5 - 5\textsuperscript{th} SMI & Data mining-based selection of solutions & Data mining algorithm \\
    \bottomrule
  \end{tabular}
  }
\end{table}

The final design is represented using rendering and visualization applications. This process is repeated in response to the feedback obtained from the client, each time combining a new set of procedures and representations to serve the purposes of creativity. Clearly, the realization of such a scenario would require standardization and interoperability between current digital platforms and applications. If the requisite support in the form of Building Information Modeling, integrated with data exchange standards, is available \cite{akin2011embedded}, such a process promises to create environments within which human creative behavior can be enhanced and extended.

\subsection{The State Space Paradox}
There have been attempts to emulate the kind of behavior we see in open systems. Genetic Algorithms, for example, that produce transformations on given genotypes are limited by the range and complexity of these symbol strings. In response to this limitation, new variations of Genetic Algorithms have been developed in which an algorithm permutes the symbol string, thus making the outcomes they induce less predictable. However, far from escaping the limitations of a closed system, this approach simply embeds one closed system (i.e., permutation of the genotypes) inside another one (i.e., generation of designs based on the genotypes). In the end, all that such a digital application can do is subsume in its state space. This is the essence of the \textit{State Space Paradox (SSP)} of computational research on creativity.

The SSP arises when an attempt is made to replicate some aspects of creative behavior by means of automated or computational closed systems. The typical argument made in systems that claim to have automated creativity is on the basis that the digital application alters the initial state space of the problem by modifying or shifting it onto another structure. For instance, Rosenman, et.al, state: “In creative design, the state space has to be [re]-formulated. This may include extending the state space of possible solutions or creating a new state space.” \cite{rosenman1993creativity}

This implies that achieving a creative solution involves the definition or redefinition of a problem space as distinct from the one(s) that were given at the outset of the digital systems operations. In other words, a closed computer system, in order to be creative, must redefine its own state space. Newell and Simon [9] (pp. 76) define a state space representation of search as the set of three indispensable components: initial state ($I$), conditions on the admissible transformations from one state to the next ($C$), and characteristics of a terminal state ($T$). Thus, the search space in a given state space of problem $\mathbf{I}$ can be defined as $S_i = \{I_i, C_i, T_i\}$.

The creative computer system, foreseen in Rosenman’s \cite{rosenman1993creativity}, and other statements that have followed its lead, then, have to be either capable of defining a new state space, say $S_j$, or be able to modify the original space, $S_i$, into a new space, $S_{i'}$. In the former case, the computer program would generate the set $\{I_j, C_j, T_j\}$; and in the latter case it would generate $\{I_{i'}, C_{i'}, T_{i'}\}$ based on the original set $\{I_i, C_i, T_i\}$. In either case, the new space is generated by the closed computer system which can only be achieved by applying $C_i$, the only operator set it has, to $I_i$, or its descendants generated by earlier applications of $C_i$. Therefore,
\begin{align*}
    \{I_j, C_j, T_j\} & \subseteq S_i \\
    \{I_{i'}, C_{i'}, T_{i'}\} & \subseteq S_i \\
\end{align*}
Thus, anything that is generated by a closed system is by definition a proper subset of its state space.

\subsection{The Consequences of the State Space Paradox}
The SSP has a serious implication for how we can see creativity-related computer systems. Tautologically, they are incapable of exhibiting the creativity we see in open systems, in a human or otherwise. This does not negate the possibility that digital creativity applications can and will invoke the SMI response in a human observer. However, they do not have the capacity to break out of their state space boundaries, regardless of the ingenuity
the programmers may have built into them.

SMI-inducing creative computer systems do not get a break when they are considered in the context of their cultural milieu. A principal reason why creativity is sought after is because it is scarce. Creativity is basically a rare human act. There are very few individuals who are considered truly creative and their lives are finite. This is a tautological outcome. If there was an overabundance of creative acts, we would no longer be willing to call them creative – or the word creative would have an entirely different meaning.

If we were able to make automated systems produce things that resemble the creative ones that humans produce, we would have an overabundance of so-called creative objects. This would, without a doubt, make us value them less, and the target we call creativity would shift. Creativity is not an absolute thing. It defies static definition and criteria of recognition. Different cultural contexts, time, place, collective agreement among individuals, and the evolution of human taste and choice, significantly influence what we call creative. Thus attaining it through well-defined and rational means will inevitably run into some form of the State Space Paradox.

\section{Conclusion and Future Directions}

While the going has been tough up to now, given the State Space Paradox, creativity-inducing or emulating digital systems have an even tougher road ahead of them. They will neither impress their creators, or anyone else for that matter, beyond the first SMI impression, nor will go beyond what we culturally consider a gimmick. This does not preclude the occasional digital application that is so smart that it will become the artists, or creators, reliable companion with its superior interface design and time-saving functions. However, in the end, a thorough analysis, beyond the SMI, will show that the human collaborator of the digital assistant will determine a product’s creativity. This is not so much a perspective of a Luddite, as it is one of cultural determinism. What we consider creative is a product of all of the traits that humans possess. For a machine to match that, would require the machine to have all traits of humans.

\bibliographystyle{abbrv}
\bibliography{mybib}

\end{document}